\documentclass[aps,12pt]{revtex4}

\usepackage{epsfig}
\usepackage{graphicx,graphics}
\usepackage{amsmath}
\usepackage{amsfonts}
\usepackage{amssymb}

\begin{document}

\title{On the appearance of traffic jams in a long chain with a shortcut in the bulk}

\author{N. Zh. Bunzarova $^{1,2}$, N. C. Pesheva $^{2}$ and J. G. Brankov$^{1}$}

\affiliation{$^1$ Bogoliubov Laboratory of Theoretical Physics, Joint Institute for Nuclear Research, \\141980 Dubna, Russia\\
$^2$ Institute of Mechanics, Bulgarian Academy of Sciences, \\1113 Sofia, Bulgaria\\
}

\begin{abstract}

The Totally Asymmetric Simple Exclusion Process (TASEP) is studied on open long chains with a shunted section between two simple chain segments in the maximum current phase. The reference case when the two branches are chosen with equal probability is considered. The conditions for the occurrence of traffic jams and their properties are investigated both within the effective rates approximation and by extensive Monte Carlo simulations for arbitrary length of the shortcut. The problem is interesting on its own because the conditions for coexistence of low- and high-density phases are essentially different from those for a simple chain between two reservoirs. Our main results are: (1) For any values of the external rates in the domain of the maximum current phase, there exists a position of the shortcut where the shunted segment is in a phase of coexistence with a completely delocalized domain wall; (2) The main features of the coexistence phase and the density profiles in the whole network are well described by the domain wall theory. Apart from the negligible inter-chain correlations, they depend only on the current through the shortcut; (3) The model displays an unexpected feature: the current through the longer shunted segment is larger than the current through the shortcut; (4) From the viewpoint of vehicular traffic, most comfortable conditions for the drivers are provided when the shortcut is shifted downstream from the position of coexistence, when both the shunted segment and the shortcut exhibit low-density lamellar flow. Most unfavorable is the opposite case of upstream shifted shortcut, when both the shunted segment and the shortcut are in a high-density phase describing congested traffic of slowly moving cars. The above results are relevant also to phenomena like crowding of molecular motors moving along twisted protofilaments.

\end{abstract}

\maketitle

\section{Introdiction}

The one-dimensional asymmetric simple-exclusion process (ASEP) is one of the simplest models of self-driven many-particle systems with particle conserving continuous time stochastic dynamics.
The process was first introduced in Ref. \cite{MGP68} as a model of kinetics of protein synthesis, i.e., of the biological process of ribosome translocation along a messenger ribonucleic acid (mRNA).
The totally asymmetric simple exclusion process (TASEP) is one of the rare examples of exactly solvable models with non-equilibrium steady states. On simple chains its stationary properties have been extensively studied and exactly solved in the thermodynamic limit for
periodic, closed and open boundary conditions, first for random sequential update and then for a number of stochastic dynamics in discrete time: forward- and backward-ordered, sublattice parallel and parallel, see the reviews \cite{D98,CSS00,S01,H01} and references therein. In the case of open chains, TASEP was shown to exhibit boundary induced phase transitions, spontaneous symmetry breaking and phase separation on the coexistence line between the low and high density phases. Recently, it has inspired a variety of modifications and extensions designed to model diverse biological problems, see, e.g., the reviews \cite{CMZ11,NKP13} and references therein.

One of the natural interpretations of TASEP is given in terms of a single-lane vehicular traffic, see the reviews \cite{CSS00,S01,H01}. Various
extensions of the basic model were devised to describe different driving conditions and drivers strategies. In this interpretation, the boundary induced first order phase transition in open chains is from free flow to congested traffic. The shock, which represents a discontinuity in the density profile, models the front of a traffic jam. The fully parallel dynamics is considered to be the most appropriate for traffic modeling and it is laid on the basis of more sophisticated update rules \cite{NS92,SSNI95,DPPP12}. The popular Nagel-Schreckenberg traffic model \cite{NS92} with maximum vehicle velocity $v_{\rm max} =1$ reduces to the TASEP with parallel update.

Obviously, the search of new control mechanisms, vehicular traffic optimization, elimination or reduction of congested traffic jams are one of the important social problems, the solution of which needs mathematical modeling. The development of computational resources provided the possibility of simulating traffic flow in discretized space and time. Much attention has been paid to cellular automaton models of traffic on roads with localized inhomogeneity modeling on- and off-ramps \cite{ PSSS01,PR02,JWW02}, intersection of two unidirectional streets \cite{BF08}, roundabout at the intersection of two streets \cite{FSS04}. Usually, traffic jams are observed at bottlenecks such as traffic accidents, lanes merging, or some of the above mentioned localized inhomogeneities. However, traffic jams often appear on crowded highways due to spontaneous fluctuations in the flow. An experimental verification that a jam can be generated in the absence of a bottleneck is reported in \cite{Setal08}. Here we confine ourselves with the stationary states of TASEP and jams which appear on the coexistence line in the phase diagram. Characteristic of these jams is that the average position of the domain wall between the low- and high-density phases is stationary, determined by the average number of particles in the system, whereas the domain wall itself performs a symmetric random walk on the chain with reflecting boundary conditions.

The special case of a network with a section of two parallel chains of equal length inserted in the bulk of a long chain was studied in our paper \cite{BPB04}. Since there are no exact results for TASEP on networks with junctions, we introduced effective injection and ejection rates for each chain segment and studied the possible phase structures of the system in terms of these rates. This approach, called later Effective Rates Approximation (ERA), turned out to be very effective in the study of the stationary phases of complex networks, composed of long linear chains, see, e.g. \cite{ PK05,NKP11,WJNW09,SMJH11, XTW11,ACH11, HA12} and the recent reviews  on TASEP over networks with complex geometry \cite{MSR12,NKP13}. We found a coexistence phase in the double-chain segment only when the head and tail chains were of equal length and in the maximum current phase. This result is interesting on its own because the set up essentially differs from the case of a simple chain coupled to reservoirs with $0< \rho_-  = 1 - \rho_+  <1/2$, when coexistence with a stationary position of the domain wall (shock) has been proved to exist. Recently, we have studied also the dependence of the phase in the double-chain segment on its position in a long but finite network \cite{PB13}. It was found that when the current through the system takes its maximum value, a simple translation of the double-chain section forward or backward along the backbone leads to a sharp change in the shape of the density profiles in the parallel chains, thus affecting the total number of particles in the network. An explanation of this phenomenon was given in terms of a finite-size dependence of the effective injection and ejection rates at the ends of the double-chain section.

Two models of TASEP on open chains with a zero-length shortcut in the bulk were introduced in
\cite{YJWHW07}: model A for molecular motor motion, and model B for vehicular traffic. In the former case the molecular motors walk along a filament which is twisted so that a motor may  jump with probability $q$ between two sites, which are far away along the filament but close in real space.
Model A was reexamined in our recent paper \cite{BPB14}. It turned out that the authors of the former work used an unfounded approximation due to which the coexistence phase in the shunted chain, in the case of maximum current through the network, was not detected. Our theoretical analysis, based on the ERA, has shown that the second (shunted) segment can exist in both low-density and high-density phases, as well as in the coexistence (shock) phase. Numerical simulations have demonstrated that the last option takes place in finite-size networks with head and tail chains of equal length, provided the injection and ejection rates at their external ends are equal and greater than one half. Then the local density distribution and the nearest-neighbor correlations in the middle chain correspond to a shock phase with completely delocalized domain wall, as is the case studied in \cite{BPB04}. Upon moving the shortcut to the head or tail of the network, the density profile takes shape typical of a high- or low-density phase, respectively, in complete parallel with the results in \cite{PB13} obtained for networks with a double-chain section.

A complicated network, consisting of an open chain of $L$ sites with a macroscopic number of zero-length shortcuts added by connecting $pL$ different pairs of sites selected randomly was considered in \cite{KSN11}. It was designed to model a randomly folded polymer chain in which there arise contacts between different polymer segments located far apart along the backbone. It was shown that the macroscopic number of shortcuts causes a drastic change in the phase diagram of the TASEP.

As we have demonstrated in \cite{BPB04,PB13}, necessary and sufficient conditions for a jam to appear in a network with a double-chain section in the middle, composed of two equivalent chains are: (i) the head and tail chain segments to be in the maximum current phase and (ii) the effective injection and ejection rates at the ends of the shunted segment to be equal. Here we will extend the above model to a family of open networks in which the number of sites in the shortcut is a free parameter, and for external injection and ejection rates which are both in the maximum current phase but may have different values. We will focus our attention on the statistical properties of the traffic jam in the shunted section, though, the distortion of the maximum-current profiles in the head and tail chains will be evaluated too. However, in the present study we will consider only the reference case when particles choose with equal probability to take the shunted segment or the shortcut. The general case of different probabilities is rather complicated and will be considered in a separate work.

The importance of the above studies rests, at least, on the following observations. First, jams are traffic phenomena with large negative impact on modern society, hence, they are in the focus of the majority of practically oriented investigations. Second, their biological significance, as noted by E. Pronina and A. B. Kolomeisky \cite{PK05}, is based on existing evidence that molecular motors kinesins move along microtubules composed of protofilaments the number of which may vary. This fact indicates the existence of junctions and other lattice defects in the microtubules which might lead to motor proteins crowding phenomena (traffic jams) that are responsible for some human diseases. Recently, traffic jams of kinesin motors transporting cargoes along microtubules have been experimentally established \cite{Leduc12}. Third, the presence of shortcuts is a quite
common feature in transportation networks. Next, the appearance of a traffic jam in the TASEP is a manifestation of a first order phase transition. It is a collective phenomenon, resulting from the directed motion and the hard-core exclusion, and, as such, has some universal features which do not depend on the details of the particle (vehicle) dynamics. Indeed, from mathematical point of view, TASEP is a discrete version of the inviscid noisy Burgers equation in the appropriate scaling limit. The analytical solutions obtained for TASEP made possible the calculation in the scaling limit of the universal critical exponents and scaling functions of the Edward-Wilkinson and Kardar-Parisi-Zhang universality classes. It was established that the versions of the TASEP, based on different update rules, belong to the same non-equilibrium finite-size scaling universality class \cite{BB05}. The critical exponents of the correlation length at the second order phase transition and of the localization length at the first order one were found to have the same value under different stochastic dynamics. Moreover, the finite-size scaling functions have the same shape for each transition order, but differ by nonuniversal prefactors which depend on the specific update.

The main aims of this work are: (a) To show that for any value of the external rates in the domain of the maximum current phase, there is a position of the shortcut such that the shunted segment is in the shock phase. Note that in \cite{BPB04,PB13} only the symmetric case of equal injection and ejection rates at the external ends of the network was considered. (b) To check on the considered  family of networks with arbitrary length of the shortcut the predicted by the Domain Wall Theory (DWT) \cite{KSKS98, SA02}  universal dependence of the main features of the shock phase in the shunted segment on the current through the shortcut, see our conjecture in \cite{BPB14}. (c) To study in detail, the statistical properties at the junction points from the viewpoint of the conditions for applicability of the DWT, and to explain the distortion of the density profiles in the head and tail segments, resulting from the presence of a shortcut.

Since the analytical results known for single chains in the thermodynamic limit cannot be a valid approximation to the properties of shortcuts with a small number of sites, extensive Monte Carlo simulations are an important tool in our investigations, along with the application of the effective rates approximation to the long segments
of the network.

The paper is organized as follows. In Section~II we define the model network, a characteristic feature of which is the existence of bifurcation and merging points between sufficiently long segments, connected by a shortcut of arbitrary length. Here we give also a brief description of our methods of analysis, provide some necessary results for the stationary properties of TASEP on a single chain in the thermodynamic limit, and formulate the basic predictions of the domain wall theory. The results of our numerical simulations for shortcuts at central position in the network are presented, in comparison with the theoretical predictions, in
Section~III. Since the applicability of the DW theory crucially depends on the properties of the junctions between the separate chain segments, the local densities at the ends of the shunted section and the shortcut, as well as the inter-chain correlations at the bifurcation and merging points of the network are studied in detail.
The cases of unequal boundary rates and off-central position of the shortcut are considered in brief in
Section~IV. The paper closes with Section~V in which we summarize our main results and conjectures.

\begin{figure}[t]
\includegraphics[width=80mm]{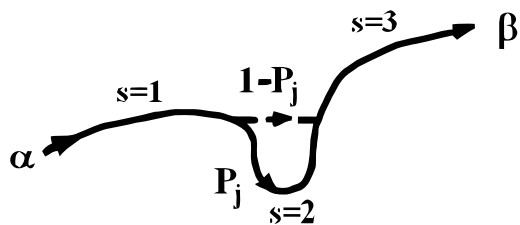} \caption{Schematic representation of a twisted filament with
a shortcut. The three long segments are labeled by $s=1,2,3$ from the left to the right and the shortcut is shown by a dashed line with an arrow. The particles are injected into the network at the left end with rate $\alpha$ and ejected from the right end with rate $\beta$; the arrows show the direction of particle hopping.}
\label{Fig1}
\end{figure}

\section{The model}

For modeling of one-dimensional directed transport of particles in biological systems we use the TASEP with stochastic dynamics in continuous time, realized in computer simulations by the so-called random-sequential update. Namely, TASEP describes a lattice gas of hard core particles on a chain, the sites of which may be empty or singly occupied, moving under the following rules: for any infinitesimal time period d$t$, just one particle attempts to hop to an empty nearest-neighbor site on the right with unit rate; if the target site is occupied, the configuration remains unchanged and another particle is randomly chosen with uniform probability distribution. In the case of an open chain, the particles are injected at the empty first site with rate $\alpha$ and removed from the occupied last site with rate $\beta$.

We model the dynamics of molecules on a twisted filament with a shortcut in the bulk, by considering TASEP on an open network consisting of 4 chains, see Fig. \ref{Fig1}. Three of the chain segments are assumed to be consecutively coupled and long enough to make reasonable the application of exact thermodynamic results. The fourth chain represents a shortcut: it allows a particle at the last site of the first segment $s=1$ to jump to the first site of the third segment $s=3$, provided the target site is empty, and has arbitrary length $L^{\rm sc} \geq 2$. At the bifurcation point the particles  try with probability $P_j$ to enter the first site of the second segment, hence, $1-P_j$ is the trial probability of entering the shortcut. Here we will confine ourselves with the reference case of $P_j = 1/2$.

The method of analysis neglects the nearest-neighbor correlations between the sites on the opposite sides of the junctions, and assumes that the chain segments are long enough, so that the exact results for the local densities at the chain ends and in the bulk hold true. Hence, for a simple chain with injection, $\alpha$, and ejection, $\beta$, rates we make use of the following asymptotic results, valid for $L \gg 1$:

(a) In the low-density phase, $\alpha < 1/2$, $\alpha < \beta$,
\begin{equation}
\rho_{bulk}=\alpha,\quad \rho_1=\alpha,\quad \rho_L= \alpha(1- \alpha)/\beta,\quad J =\alpha(1- \alpha);
\label{LDas}
\end{equation}

(b) In the high-density phase, $\beta < 1/2$, $\beta < \alpha$,
\begin{equation}
\rho_{bulk}=1 -\beta,\quad \rho_1= 1- \beta(1- \beta)/\alpha,\quad \rho_L= 1- \beta,\quad J =\beta(1- \beta);
\label{HDas}
\end{equation}

(c) In the maximum current phase, $\alpha > 1/2$ and $\beta > 1/2$,
\begin{equation}
\rho_{bulk}=1/2,\quad \rho_1= 1- 1/(4\alpha),\quad \rho_L= 1/(4\beta),\quad J = 1/4;
\label{MCas}
\end{equation}

(d) On the coexistence line, $\alpha = \beta < 1/2$, in the presence of a delocalized domain wall (for open systems), the local density profile changes linearly from $\rho_1 = \alpha$ to $\rho_L = 1- \alpha$, and the nearest-neighbor correlations have a parabolic shape with maximum value
\begin{equation}
\max_x F_{\rm cor}(x) = F_{\rm cor}(L/2)= (1- 2\alpha)^2/4.
\label{maxFcor}
\end{equation}

The stationary state of each segment is determined by the effective input, $\alpha_s^*$, and output, $\beta_s^*$, rates for each segment $s=1,2,3$ which are defined by the following rules:
\begin{equation}
J^{(s)}= \beta_s^* \rho_L^{(s)} = \alpha_s^*\left(1- \rho_1^{(s)}\right).
\label{eff}
\end{equation}
Here $J^{(s)}$ is the current through the segment $s$, $\rho_1^{(s)}$ and $\rho_L^{(s)}$ are  the average local densities at the first and last site $L$ of that segment; obviously, $\alpha_1^* =\alpha$, $\beta_3^* =\beta$. The equalities in (\ref{eff}) neglect the inter-chain correlations.

The continuity of the current leads to a set of equations between these quantities. The solutions for $\alpha_s^*$ and $\beta_s^*$ determine the nonequilibrium stationary phase of each segment $s=1,2,3$, and the shortcut itself, provided the latter is long enough.

To model the average density profile in the long segments, one may use the phenomenological domain wall (DW) theory \cite{KSKS98,SA02}. Its main idea is that, when $\alpha \not= 1- \beta$, each reservoir tends to enforce a domain in the chain with its own density: the left domain with density  $\rho_- = \alpha$, and the right one with density $\rho_+ = 1- \beta$. At a given time $t$ these domains may coexist, being separated by a domain wall. The stochastic dynamics implies that the domain wall performs a random walk on the chain. On the coexistence line $\alpha = \beta < 1/2$ the random walk is symmetric, with diffusion constant $D = \alpha(1-\alpha)/(1-2\alpha)$ and reflecting boundary conditions.

The question if the DW theory is applicable to the central segment of our network is not trivial and will be studied here in some detail by comparison with results of computer simulations. The problem is that the ends of the shunted chain are not connected to reservoirs of particles with appropriate low and high density. Instead, its left end is coupled to the right end of the first segment where the local density is $\rho_L^{(1)} = 1/(4\beta_1^*) < 1/2$, and the right end is coupled to the left end of the third segment, where the local density is $\rho_1^{(3)} = 1 - 1/(4\alpha_3^*) > 1/2$. These conditions differ from the connections to reservoirs in that there can be significant correlations at the junctions between the segments which are absent in the case of reservoirs. Thus, the very possibility of finding a shunted chain in the coexistence phase under the above boundary conditions is of its own interest.

The predictions of the DW theory for the endpoints of the shunted segment read:
\begin{equation}
\rho_{1,L}^{(2)} = \rho_{\mp}^{(2)} \equiv   1/2 \mp \sqrt{J^{\rm sc}},
\label{ends2}
\end{equation}
and for the maximum value of the nearest-neighbor correlations:
\begin{equation}
\max_x F_{\rm cor}(x) = \left[\rho_{+}^{(2)}- \rho_{-}^{(2)}\right]^2 = 1/4 -J^{(2)} = J^{\rm sc},
\label{maxcor}
\end{equation}
where
\begin{equation}
\rho_{\pm}^{(2)}= \rho_{\pm}^{(2)}\left(J^{(2)}\right), \quad \rho_{\pm}(J):= \left(1\pm \sqrt{1-4J}\right)/2.
\label{ropm}
\end{equation}

The main aim of the study is to analyze the conditions under which the shunted segment is found in the coexistence phase. Next, we will check if the main features of the coexistence phase in that segment agree with the predictions of the DW theory given by Eqs. (\ref{ends2}) and (\ref{maxcor}), which imply that these features depend only on the current through the shortcut, and not on its structure.

\section{Central position of the shortcut}

For our reference computer simulations we have taken $L_1 =L_2 =L_3 = 400$ sites, which is about the evaluated in \cite{Leduc12} number of steps taken by a molecular motor during its average residency time on a microtubule. The length of the shortcut $L^{\rm sc}$ varies from 2 to 100 sites. The numerical data was averaged over a sample of 100 independent runs of length $2^{24}$ trials each, which amounts to more than $10^6$ Monte Carlo steps per site. The statistical error was estimated by comparing the above data to the one obtained by averaging over a twice larger sample consisting of 200 independent runs of length $2^{24}$ trials each. The so estimated relative statistical error in the currents is less than $2\times 10^{-4}$, and in the local densities at the endpoints of the chain segments is less than 2\% for all the studied shortcut lengths.

The finite-size shift, estimated from the comparison of the measured total current $J^{\rm sim}\simeq 0.25091$ with the theoretical one $J_{\rm max} =1/4$ yields a relative value slightly less than 0.4\%. More important, the density profile in the bulk of a finite chain in the MC phase is not exactly constant, as in the thermodynamic limit $\rho_{bulk}=1/2$, but has a slight slope downwards in the direction of hopping. This seems to be the reason why the effective injection and ejection rates of the middle segment vary upon shifting its position in the network \cite{PB13}.
\begin{figure}[t]
\includegraphics[width=100mm]{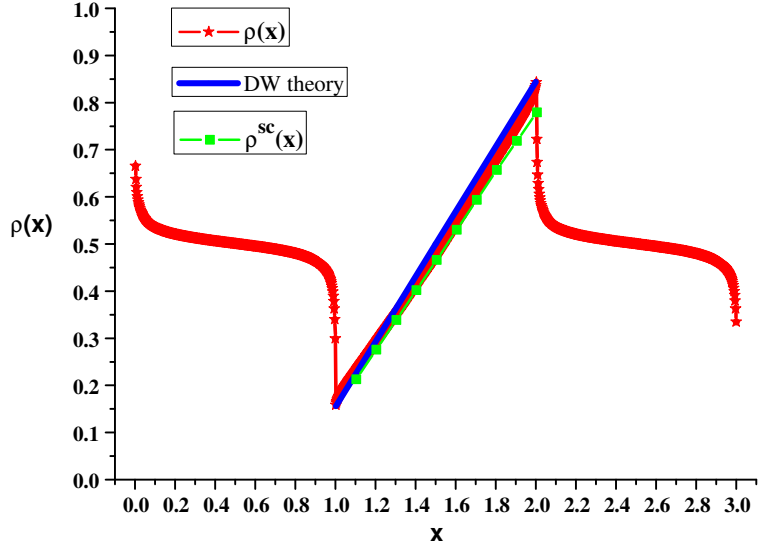} \caption{(Color online) Simulation results for the local density profile in the whole network. The local density profiles in the second segment (red stars) and in the shortcut (green circles) are compared to the predicted by the DW theory linear dependence (blue line) for a shortcut of length 10 sites.}
\label{Fig2}
\end{figure}

Here we consider the case when the head and tail segments have an equal length $L_1 =L_3$. To estimate the conditions for appearance of a domain wall in the shunted segment $s=2$, we use arguments based on the ERA. Thus, within the first approximation we neglect the nearest-neighbor correlations between the different chains. By choosing $\alpha >1/2$ and $\beta >1/2$ we ensure that the whole network carries a maximum current of particles $J=1/4$ and so do the head and tail segments, $s=1$ and $s=3$, respectively. Since these segments are in the MC phase, we have the following particle densities at their ends:
\begin{equation}
\rho_1^{(1)}= 1- 1/(4\alpha),\quad \rho_L^{(1)}= 1/(4\beta_1^*),\quad \rho_1^{(3)}= 1- 1/(4\alpha_3^*),\quad \rho_L^{(1)}= 1/(4\beta).
\label{ends13}
\end{equation}
Here the upper index of the average density labels the segment, and the lower one the site in the segment: lower index 1 stays for the first site and $L$ for the last one;  $\beta_1^*$ is the effective ejection rate from the first segment, $\alpha_3^*$ is the effective injection rate in the third segment.

A single chain is on the coexistence line (domain-wall phase) when the input and output rates satisfy the relation $\alpha = \beta \leq 1/2$. As $s=1$ and $s=3$ are of equal length, and both in the same MC phase, one could guess that if the correlations at the junctions with the inner chains are neglected, the profiles in these chains should be identical (congruent), with $\beta_1^*= \beta$ and $\alpha_3^* =\alpha$. However, it turns out that the very presence of a shortcut in the bulk distorts the density profiles of the segments $s=1$ and $s=3$ in an asymmetric way. In any case, we shall interpret the last site of segment $s=1$ as a low-density $\rho_L^{(1)} < 1/2$ reservoir, and the first site of segment $s=3$ as a reservoir with high-density $\rho_1^{(3)} >1/2$.  Since in our model the probability of choosing any of the chains in the double-chain section equals to 1/2, and there are no rules of preference at the merging point, the condition for phase coexistence in the middle section is expected to be $\rho_L^{(1)}/2 \simeq 1- \rho_1^{(3)}$, provided the shortcut is long enough. As we will show below, this condition is met when $\alpha =\beta >1/2$, if the shunted segment $s=2$ is at central position in the network.
\begin{figure}[t]
\includegraphics[width=100mm]{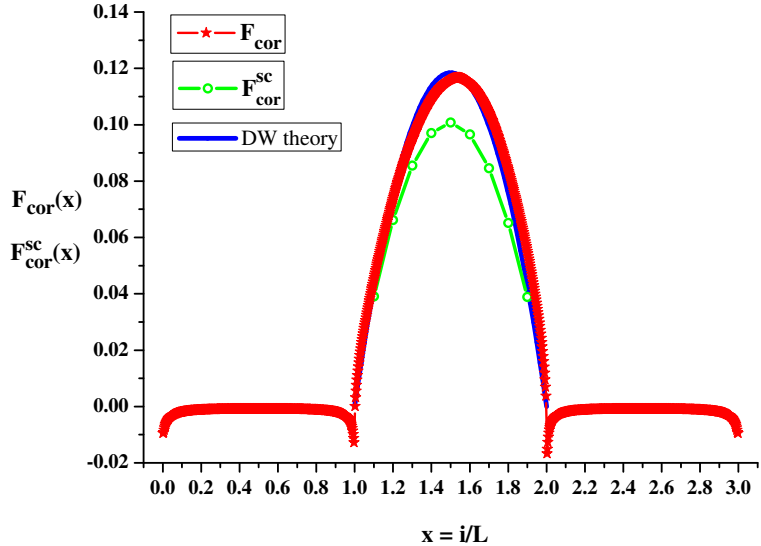} \caption{(Color online) Simulation results for the nearest-neighbor correlations in the whole network. Those for the second segment (red stars) and the shortcut (green circles) are compared to the predicted by the DW theory parabolic shape (blue line) for a shortcut of length 10 sites.}
\label{Fig3}
\end{figure}

Here we present results of our computer simulations at the point $(\alpha,\beta)= (0.75,0.75)$ on the diagonal in the domain of the maximum current phase of a single chain. In all the figures in this section, when data for the main segments
are plotted, the abscissa is the normalized coordinate $x=i/L$, $i=1,2, \dots, 3L$ ranging in the interval $(0,3]$. However, when data for the shortcut are plotted, for convenience of comparison with the shunted segment $s=2$, the abscissa is stretched to $x=1 +i/L^{\rm sc}$, $i=1,2,\dots, L^{\rm sc}$ so that its range is $(1,2]$.

\subsection{Properties of the shunted segment}

The numerical data for a shortcut with $L^{\rm sc} =10$, illustrated by Fig. \ref{Fig2}, exhibit unexpectedly good agreement with the DW theory prediction for the density at the endpoints of the shunted segment: $\rho_1^{(2)} \simeq 0.158$, to be compared with $\rho_{-} \simeq 0.156$, and
$\rho_L^{(2)} \simeq 0.836$, to be compared with $\rho_{+} \simeq 0.844$.
The values of $\rho_{\pm}$ are calculated according to Eq. (\ref{ends2}) with the evaluated value of
$J^{\rm sc}\simeq 0.118$.

The nearly perfect parabolic shape of the nearest-neighbor correlations in the second segment, see Fig. \ref{Fig3}, is described by the best fit quadratic approximation
$F_{\rm cor}^{\rm fit}(x) =A + B x + C x^2$, with $A=-0.94744$, $B=1.42116$, and $C=-0.47372$,
which yields $\max_x F_{\rm cor}^{\rm fit}(x) = 0.11843$, in very good agreement with the numerically evaluated $\max_x F_{\rm cor}(x) \simeq 0.117$, and
in excellent agreement with the value of $J^{\rm sc}\simeq 0.118$.
\begin{figure}[t]
\includegraphics[width=100mm]{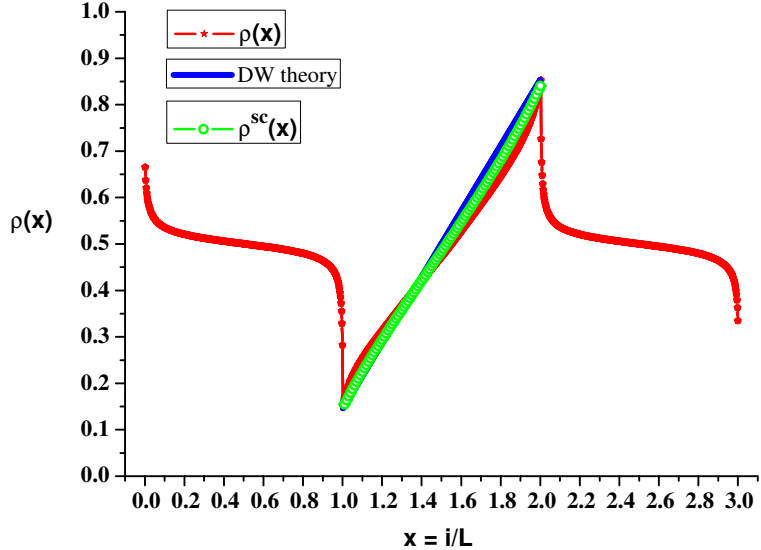} \caption{(Color online) Simulation results for the local density profile in the whole network. The local density profiles in the second segment (red stars) and in the shortcut (green circles) are compared to the predicted by the DW theory linear dependence (blue line) for a shortcut of length 100 sites.}
\label{Fig4}
\end{figure}

Next we present analogous results for a rather long shortcut containing 100 sites.
In this case, the local density profile in both the shunted second segment and the shortcut indicates
a coexistence phase in each of them, see Fig. \ref{Fig4}. However, in this case there is a noticeable
deviation from the linear dependence on the normalized coordinate $x=i/L$, $i=L+1,\dots, 2L$.
The corresponding numerical results are: for the left ends $\rho_1^{(2)} \simeq 0.153$ and $\rho_1^{\rm sc} \simeq 0.157$, to be compared to the DW prediction $\rho_{-} \simeq 0.1464$; for the right ends $\rho_L^{(2)} \simeq 0.846$ and $\rho_{100}^{\rm sc} \simeq 0.843$, to be compared to $\rho_{+} \simeq 0.8536$.
Here the values of $\rho_{\pm}$ are calculated from Eq. (\ref{ends2}) with the evaluated value of
$J^{\rm sc}\simeq 0.12502$. The agreement is rather good, although there is a noticeable deviation
from the DW prediction, especially for the local density at the left end of the shunted segment.
\begin{figure}[t]
\includegraphics[width=100mm]{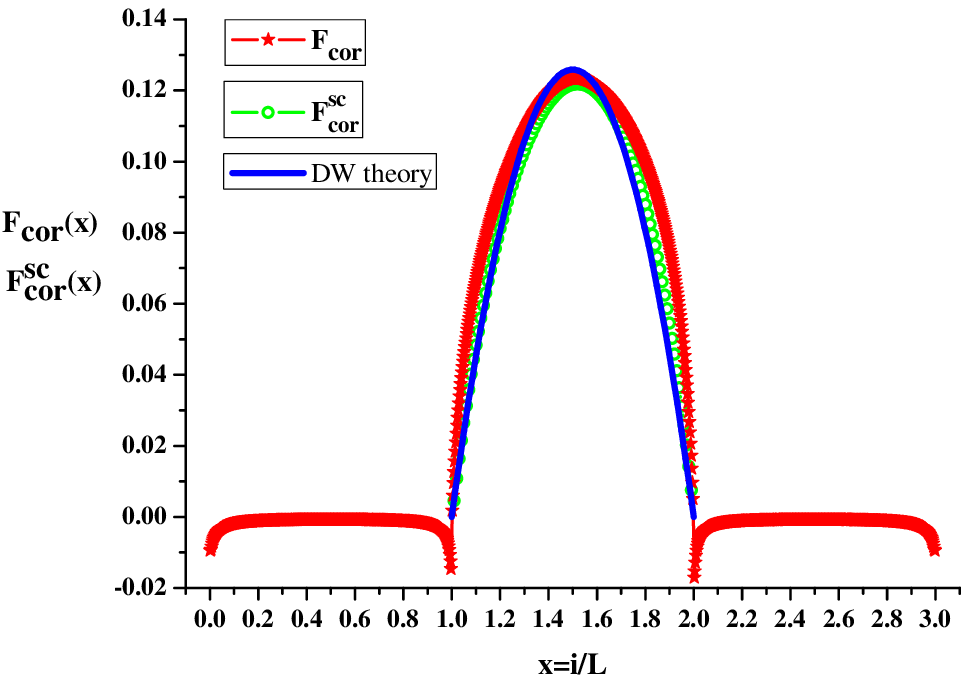} \caption{(Color online) Simulation results for the nearest-neighbor correlations in the whole network. Those for the second segment (red stars) and the shortcut (green circles) are compared to the predicted by the DW theory parabolic shape (blue discs) for a shortcut of length 100 sites.}
\label{Fig5}
\end{figure}

Small deviations from the parabolic shape is observed also in the plot of the nearest-neighbor correlations,
both in the shunted segment and in the shortcut of length 100 sites, see Fig. \ref{Fig5}.
The best quadratic fit for the shunted segment is
$F_{\rm cor}^{\rm fit}(x) =A + B x + C x^2$, with $A=-1.00711$, $B=1.51067$, and $C=-0.50356$,
which yields $\max_x F_{\rm cor}^{\rm fit}(x) = 0.125885$,
in very good agreement with the value of $J^{\rm sc}\simeq 0.12502$. On the other hand, the numerical evaluation of the maxima in the shunted segment and in the shortcut provides
$\max_x F_{\rm cor}(x) \simeq 0.1234$ and $\max_x F^{\rm sc}_{\rm cor}(x)\simeq 0.1219$,
which deviate from the DW theory prediction within 1.3~\% in the former case and 2.6~\% in the latter one. In the symmetric case of $L^{\rm sc} =L=400$,
the results are almost the same, $\max_x F_{\rm cor}(x) = \max_x F^{\rm sc}_{\rm cor}(x)\simeq 0.1239$,
which differs from the DW theory prediction by less than 1.3~\%.
\begin{figure}[b]
\includegraphics[width=90mm]{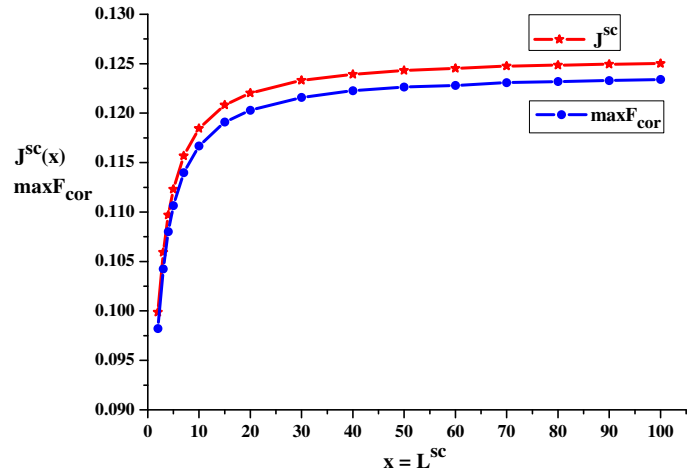} \caption{(Color online) Simulation results for the maximum value of the nearest-neighbor correlations on the shunted segment, $\max_x F_{\rm cor}(x)$ (blue discs),  compared to the predicted by the DW theory value $J^{\rm sc}$ (red stars) for shortcuts of different length.}   \label{Fig6}
\end{figure}

The agreement with the prediction (\ref{maxcor}) of the DW theory for the maximum value of the nearest-neighbor correlations on the shunted segment is illustrated in Fig. \ref{Fig6}. The discrepancy is less than 2~\% in the whole range of shortcut lengths.

Summarizing the above results, on the one side we see very good agreement with the DW picture, and the other side persistent small deviations. The latter can be due to both finite-size effects and to small but noticeable
inter-chain correlations.

\subsection{Junctions and inter-chain correlations}

\begin{figure}[b]
\includegraphics[width=100mm]{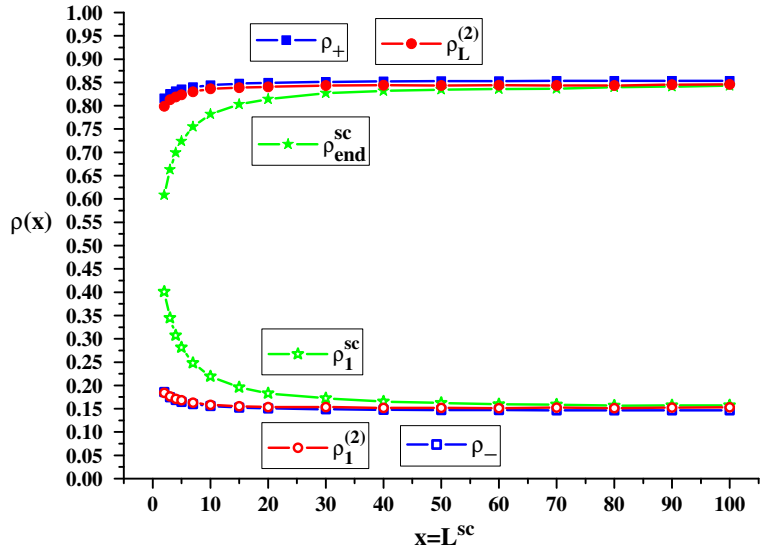} \caption{(Color online) Simulation results for the local density at the first site of the shunted segment, $\rho_1^{(2)}$ (red circles), and the shortcut, $\rho_{1}^{\rm sc}$ (green empty stars), as well as for the the last site of the shunted segment, $\rho_L^{(2)}$ (red filled circles), and the shortcut, $\rho_{\rm end}^{\rm sc}$ (green filled stars), as a function of the shortcut length. For comparison, the predicted by the DW theory values for the shunted segment are shown: $\rho_{-}^{(2)}$ (blue empty squares) for the first site and $\rho_{+}^{(2)}$ (blue filled squares) for the last site.}   \label{Fig7}
\end{figure}

In the general case, the behavior of the numerically evaluated and predicted local densities (\ref{ends2}) at the end points of both the second segment and the shortcut are shown in Fig. \ref{Fig7}, as a function of the shortcut length $L^{\rm sc}$ in the range from 2 to 100 sites.

Note that for all shortcut lengths the relative deviation of $\rho_1^{(2)}$ from $\rho_{-}^{(2)}$ is within 3~\%, and the relative deviation of $\rho_L^{(2)}$ from $\rho_{+}^{(2)}$ is less than 2~\%.
The local densities at the first and last site of the shortcut strongly deviate from the DW predictions for small $L^{\rm sc}$ but steadily approach the latter values with the growth of the shortcut length. It is seen that for $L^{\rm sc}\ge 100$ the shortcut becomes nearly equivalent to the shunted segment.
Possible explanation of the observed small deviations will be discussed in the remainder of this section.

\begin{figure}[t]
\includegraphics[width=100mm]{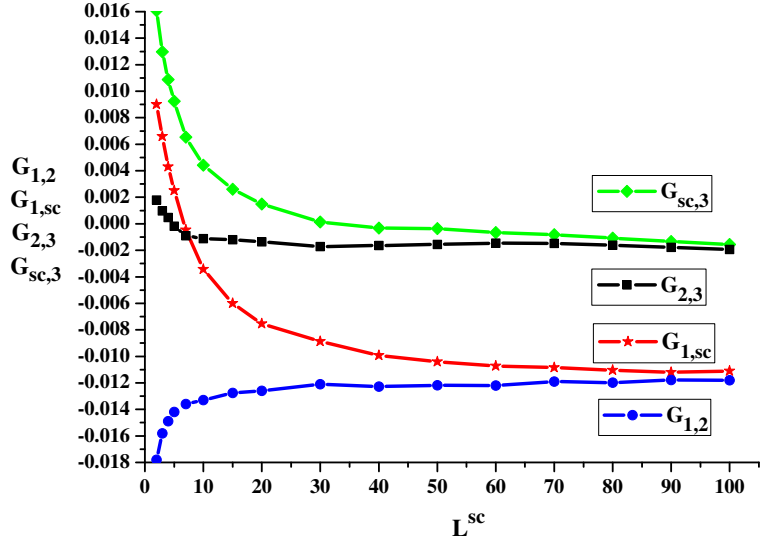} \caption{(Color online) Simulation results for the inter-chain correlations in dependance on the length of the shunted segment (see text).}   \label{Fig8}
\end{figure}

Under the random sequential update, the inter-chain correlations measure the difference between the actual current of particles and its mean-field approximation. Taking into account that in our model there is a rule conducting the currents out of the bifurcation
point with probability $P_j$ to the shunted segment and with the remaining $1-P_j$ to the shortcut, we obtain
\begin{equation}
G_{1,2} =-\frac{1}{P_j}J^{(2)}+ \rho_L^{(1)}\left(1-\rho_1^{(2)}\right), \quad
G_{1,\rm sc} =-\frac{1}{1-P_j}J^{\rm sc}+ \rho_L^{(1)}\left(1-\rho_1^{\rm sc}\right), \label{G12}
\end{equation}
where $G_{1,2}$ denotes  the nearest-neighbor correlation between the first segment
and the second one, and $G_{1,\rm sc}$ between the first segment and the shortcut. In our case $P_j =1/2$.
The behavior of these inter-chain correlations, as well as the inter-chain correlations between the second and
third segments, $G_{2,3}$, and between the shortcut and the third segment, $G_{\rm sc,3}$, is shown in
Fig.~\ref{Fig8} as a function of the shortcut length.
One sees that the correlations between the shunted segment and the third one are negligible in the whole range studied. The correlations between the shortcut and the third segment are positive and significant only for small $L^{\rm sc}$ up to 20, then quickly drop down to absolute values within the statistical error. Hence, the DW approximations $\rho_{\mp}^{(2)}$ are fairly close to the actual values of $\rho_{1,\rm end}^{\rm sc}$ for long enough shortcuts, say, for $L^{\rm sc} \geq 50$. On the other hand, the correlations between the head and shunted segment, as well as those between the head segment and the shortcut, tend to a common value of about $-0.0125$ as $L^{\rm sc}$ increases.

The negative values of the inter-chain correlations $G_{1,2}$ and $G_{1,\rm sc}$, see Fig. \ref{Fig8}, imply that the actual currents through the shunted segment and the shortcut exceed the corresponding mean field approximations at the bifurcation point:
\begin{equation}
J^{(2)} = P_j \left[\rho_L^{(1)}\left(1-\rho_1^{(2)}\right)-G_{1,2}\right], \quad
J^{\rm sc} =(1-P_j)\left[\rho_L^{(1)}\left(1-\rho_1^{\rm sc}\right)-G_{1,\rm sc}\right]. \label{J21sc}
\end{equation}

\begin{figure}[t]
\includegraphics[width=100mm]{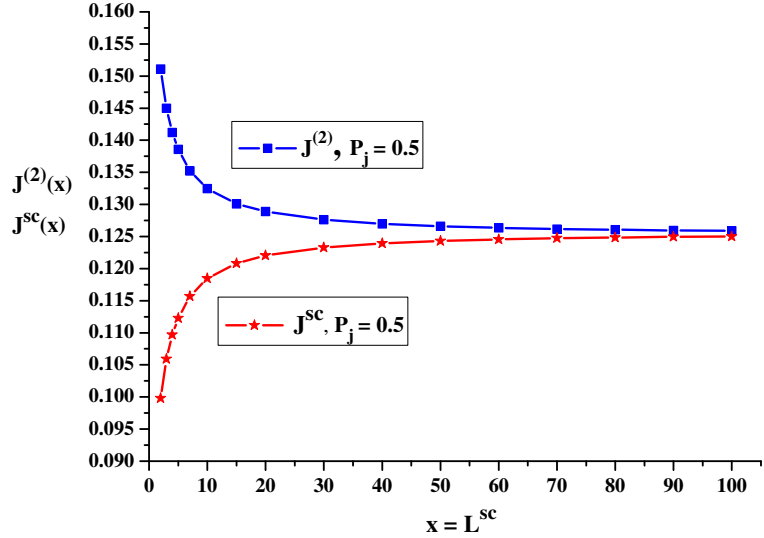} \caption{(Color online) Dependance of the currents through the shunted segment (blue squares) and the shortcut (red stars) on the length of the shortcut.}  \label{Fig9}
\end{figure}

Turning back to Fig. \ref{Fig7}, we note that $\rho_1^{\rm sc}> \rho_1^{(2)}$ for all values $L^{\rm sc}< L$.
In view of Eqs. (\ref{G12}) and the constraint $J^{\rm sc}+J^{(2)}=1/4$, this implies the following inequalities
\begin{equation}
J^{(2)} > \frac{1}{8} + \frac{1}{4}\left(G_{1,\rm sc}-G_{1,2}\right), \quad
J^{\rm sc} <  \frac{1}{8} - \frac{1}{4}\left(G_{1,\rm sc}-G_{1,2}\right),\quad L^{\rm sc}< L. \label{ineqJ}
\end{equation}
Since $G_{1,\rm sc} > G_{1,2}$ for all $L^{\rm sc}< L$, it follows that $J^{(2)} > 1/8 > J^{\rm sc}$, i.e.,
it turns out that the current through the longer shunted segment is larger than the current through the shortcut. This observation is illustrated in Fig.~\ref{Fig9}.

Evidently, at $L^{\rm sc}= L$ and $P_j =1/2$, the shunted segment and the shortcut become equivalent, hence,
$\rho_1^{\rm sc}= \rho_1^{(2)}$, $G_{1,\rm sc} = G_{1,2}$  and $J^{(2)} = J^{\rm sc} = 1/8$. Finally, for $L^{\rm sc}> L$ the shunted segment and the shortcut exchange their places.

At the merging point there are no rules other than the simple exclusion and we have
\begin{equation}
G_{2,3} =\rho_L^{(2)}\left(1-\rho_1^{(3)}\right)- J^{(2)}, \quad
G_{{\rm sc},3} =\rho_{\rm end}^{\rm sc}\left(1-\rho_1^{(3)}\right) - J^{\rm sc},\label{G23}
\end{equation}
where $G_{2,3}$ denotes  the nearest-neighbor correlation between the second segment
and the third, and $G_{{\rm sc},3}$ between the shortcut and the third segment.

\subsection{Density profiles in the head and tail segments}

As already mentioned, a significant asymmetric distortion of the density profiles in segments $s=1$ and $s=3$
occurs due to the very presence of the shortcut in the bulk. First, we note that the local densities at the
external ends of the network fluctuate within statistical error around $\rho_1^{(1)} \simeq 0.6654$ and
$\rho_L^{(3)} \simeq 0.3345$, independently of the length of the shortcut $2\leq L^{\rm sc} \leq L=400$.
In the studied case of $\alpha =\beta =0.75$, the deviation of the evaluated $\rho_1^{(1)}$ from the theoretical value $\rho_1^{(1)}= 1- 1/(4\alpha)= 0.666(6)$, and of $\rho_L^{(3)}$ from $\rho_L^{(3)}= 1/(4\beta)= 0.333(3)$ is definitely due to the finite size of the system.

To approach the local densities at the inner ends of the head and tail segments, we use two exact expressions for the current in the shunted segment,$J^{(2)}$, at $P_j = 1/2$:
\begin{equation}
J^{(2)}=\frac{1}{2}\left[\rho_L^{(1)}(1- \rho_1^{(2)}) - G_{1,2}\right] =
\rho_L^{(2)}(1- \rho_1^{(3)}) - G_{2,3}. \label{J2G}
\end{equation}
From the above equalities we obtain the local densities of interest in the form
\begin{equation}
\rho_L^{(1)} = \frac{2J^{(2)} + G_{1,2}}{1- \rho_1^{(2)}} ,
\qquad \rho_1^{(3)} = 1- \frac{J^{(2)} + G_{2,3}}{\rho_L^{(2)}}. \label{L113}
\end{equation}

\begin{figure}[t]
\includegraphics[width=100mm]{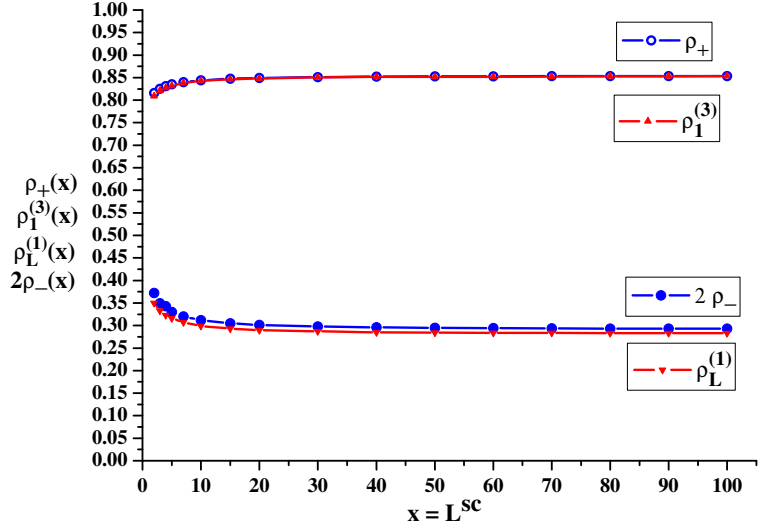} \caption{(Color online) The local densities at the end of the head chain $\rho_L^{(1)}$ (red down triangles), and the first site of the tail chain $\rho_1^{(3)}$ (red up triangles), compared to the approximations given in Eq. (\ref{InnEnds}) (filled and empty blue circles, respectively) based on the DW theory and the neglect of inter-chain correlations.}   \label{Fig10}
\end{figure}

As we have already shown, the shunted segment is in the coexistence phase with $\rho_{1,L}^{(2)}$
nicely obeying the DW prediction (\ref{ends2}). Taking into account that $\rho_{-}^{(2)}\rho_{+}^{(2)} = J^{(2)}$, we obtain
\begin{equation}
\rho_L^{(1)} \simeq 2\rho_{-}^{(2)}\left[1 + \frac{G_{1,2}}{2J^{(2)}}\right],\quad
\rho_1^{(3)} \simeq 1- \rho_{-}^{(2)}\left[1 + \frac{G_{2,3}}{J^{(2)}}\right]. \label{G12nn}
\end{equation}
From Fig. \ref{Fig8} we see that $G_{1,2}$ is negative and approaches -0.0125 as $L^{\rm sc}\rightarrow L=400$,
while $|G_{2,3}|< 0.02$ for all $L^{\rm sc}$.
Therefore, within 5\% error in the value of $\rho_L^{(1)}$ and 0.4 \% error in $\rho_1^{(3)}$, we can neglect the contribution of the inter-chain correlations in Eqs. (\ref{G12nn}) and obtain the very simple approximations
\begin{equation}
\rho_L^{(1)} \simeq 2\rho_{-}^{(2)},\qquad
\rho_1^{(3)} \simeq \rho_{+}^{(2)}. \label{InnEnds}
\end{equation}
The excellent agreement of expressions (\ref{InnEnds})  with the computer simulations data is illustrated in
Fig.~\ref{Fig10}.

Now we take into account that the total density $\rho_L^{(1)} \simeq 2\rho_{-}^{(2)}$ of the left-hand reservoir is shared as follows: a fraction of $P_j$ is available to the shunted segment $s=2$, the remaining fraction of $1-P_j$ being available to the shortcut. Therefore, at $P_j =1/2$ the inner end of the head segment act as two independent reservoirs of particles with equal effective density $\tilde{\rho}_L^{(1)} \equiv \rho_L^{(1)}/2 \simeq \rho_{-}^{(2)}$, coupled to each of the two branches in the double-chain section. Since at the merging point there are no preferences for the entrance of
particles into the tail segment $s=3$, the right-hand reservoir has the same density $\rho_1^{(3)} \simeq \rho_{+}^{(2)} = 1- \rho_{-}^{(2)}>1/2$ for both the shunted segment and the shortcut. Thus, by neglecting the small effects of the inter-chain correlations, we arrive at the condition for phase coexistence in both branches of the double-chain section, treated as independent simple chains coupled to the appropriate reservoirs.

To quantitatively evaluate the distortions in the maximum-current profiles in the head and tail segments, we
note that the current $J^{\rm sc}$ monotonically increases with $L^{\rm sc}$, from $J^{\rm sc} \simeq 0.09987$ at $L^{\rm sc}=2$, to $J^{\rm sc} \simeq 0.12546$ at $L^{\rm sc}=L=400$. Accordingly, see Eq. (\ref{ends2}),
$\rho_{-}^{(2)}$ decreases from $\rho_{-}^{(2)}\simeq 0.18398$ to $\rho_{-}^{(2)}\simeq 0.14580$, while
$\rho_{+}^{(2)}$ increases from $\rho_{-}^{(2)}\simeq 0.81602$ to $\rho_{-}^{(2)}\simeq 0.85420$. From our data
we readily conclude that:

(i) At $L^{\rm sc}=2,3,4$, the right end of the head profile is \textit{shortened} by at most
$2\rho_{-}^{(2)}-1/3 \simeq 0.0346$, while for $L^{\rm sc}\geq 5$ it is  \textit{extended downward} by at most
$1/3 - 2\rho_{-}^{(2)}\simeq 0.04173$.

(ii) For all $L^{\rm sc}=2,3,\dots ,400$, the left end of the tail profile is \textit{extended upward} at least by $\rho_{+}^{(2)}-2/3 \simeq 0.14935$ at $L^{\rm sc}=2$, and at most by $\rho_{+}^{(2)}-2/3\simeq
0.1875$ at $L^{\rm sc}=400$.

(iii) The small negative inter-chain correlations lead to a slight enhancement of the above effects: further small decrease in $\rho_L^{(1)}$ and increase $\rho_1^{(3)}$, as is seen from Eqs. (\ref{G12nn}).

These results have natural physical implications: the presence of a double-chain section in the bulk, with the
present traffic rules at the bifurcation point, leads to depletion of particles at the end of the
head tail (for $L^{\rm sc}\geq 5$) and their excessive accumulation at the first site of the tail segment.

\section{Off-central position of the shortcut}

\begin{figure}[b]
\includegraphics[width=100mm]{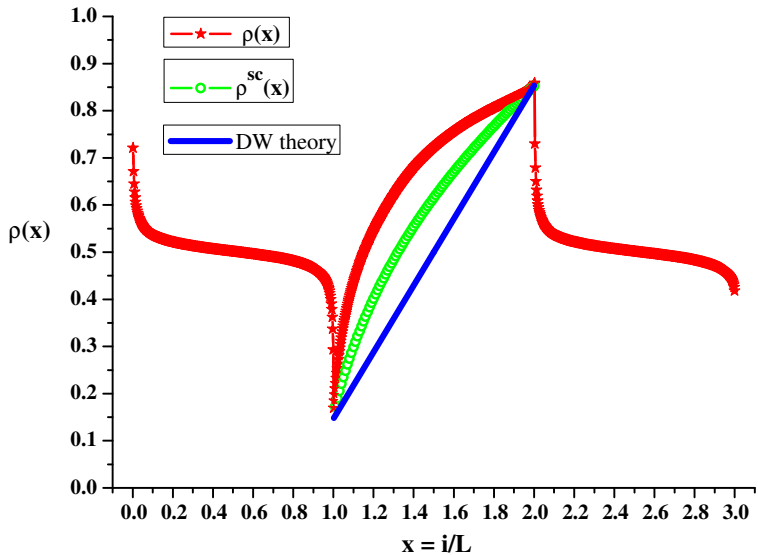} \caption{(Color online) Simulation results for the local density profile in the network under central position of the shortcut,$L_1 = L_3$, when the point of the external rates is shifted off-diagonally to $(\alpha,\beta) = (0.9,0.6)$. The local density profiles in the shunted segment (red stars) and the shortcut of length $L^{\rm sc}=100$ (green circles) are compared to the linear profile predicted by the DW theory (blue line). }   \label{Fig11}
\end{figure}

In the case of long but finite chains, moving the point of external rates $(\alpha,\beta)$ in the domain of the maximum current phase away from the diagonal $\alpha =\beta >1/2$  changes the effective rates $\alpha_2^*$ and $\beta_2^*$ for the shunted segment and brings it out of the coexistence phase. We numerically studied the particular case of $(\alpha,\beta) = (0.9,0.6)$ when one expects the balance between $\alpha_2^*$ and $\beta_2^*$ to be broken to $\alpha_2^*> \beta_2^*$. Indeed, computer simulations show that in this case the density profiles of both the shunted segment and the shortcut of length $L^{\rm sc}=100$ bend upward to resemble the shape of the high density phase, see Fig. \ref{Fig11}. However, the balance $\alpha_2^* \simeq \beta_2^*$ is recovered upon slight shift backward of the position of the shortcut to $L_1 = 410$, $L_3 = 390$, when computer simulations show that the density profiles of both the shunted segment and the shortcut with length $L^{\rm sc}=100$ are almost linear, which indicates the coexistence phase in both.

\begin{figure}[t]
\includegraphics[width=100mm]{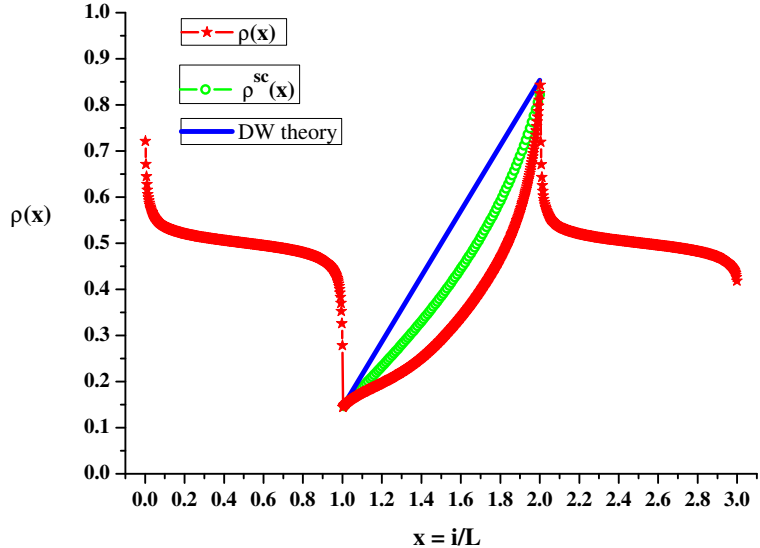} \caption{(Color online) Simulation results for the local density profiles in the shunted segment (red stars) and the shortcut of length $L^{\rm sc}=100$ (green circles) in the network with $(\alpha,\beta) = (0.9,0.6)$, under excessively shifted position of the shortcut to $L_1 = 420$, $L_3 = 380$. For comparison the linear profile predicted by the DW theory is shown by the blue line.}   \label{Fig12}
\end{figure}

Expectedly, when the shortcut is shifted further backward to the position $L_1 = 420$, $L_3 = 380$, the balance $\alpha_2^* \simeq \beta_2^*$ changes in the opposite direction to  $\alpha_2^* < \beta_2^*$. Indeed, computer simulations show that now the density profiles of both the shunted segment and the shortcut bend downward, similarly to the case of a low density phase, see Fig. \ref{Fig12}.

Thus, we conjecture that depending on the position of the shortcut, the density profile in the whole network may have different shapes, typical for the stationary phase structures
MC-HD-MC, MC-DW-MC, and MC-LD-MC, where DW stays for the domain wall (coexistence) phase. In the first two cases, on crossing the boundary of the HD phase the average particle velocity will decrease exponentially fast in conformity with the local density increase. In the thermodynamic limit this takes place in the so called boundary layer. At the boundaries of the maximum current phase the particles accelerate following a power-law with the distance, which is a consequence of the fact that the local density approaches the bulk value $\rho =1/2$ approaches its bulk value of $\rho =1/2$ from above as $i^{-1/2}$, $i =1,2,3,\dots$, $i\ll L$, near the left boundary and from below as $(L-i)^{-1/2}$, $L-i =1,2,3,\dots$, $L-i\ll L$, near the right boundary.

\section{Discussion}

Naturally, simple models as the one studied here are not expected to give quantitative results which favorably compare to real world observations. Nevertheless, the universality hypothesis implies that results obtained from the solution of the simplest TASEP models apply to a much wider collection of stochastic systems with different short-ranged inter-particle interactions and microscopic dynamics. Thus, these results may play an important role in our better understanding of the rich and in many instances unexpected phenomena exhibited by non-equilibrium systems. Some of these phenomena have no analogue in the equilibrium. Let us just mention the appearance of boundary induced phase transitions in one dimension, the observed position-induced phase change in a double-chain section, or in the shunted section only, when the network is in the maximum current regime, as well as the appearance of a shock phase in the shunted segment. Using simple models to study different effects and their interplay has proven a very powerful approach in equilibrium statistical mechanics and now the same approach is widely employed to non-equilibrium systems.

In the considered case of $P_j=1/2$, our extensive Monte Carlo simulations lead us to the following main results concerning the appearance and properties of traffic jams in a sufficiently long shunted segment between head and tail chains carrying maximum current:

1.  For any values of the external rates in the domain of the maximum current phase, $\alpha >1/2$  and $\beta >1/2$, there exists a position of the shortcut where the shunted segment is in a phase of coexistence with a completely delocalized domain wall. The longer is the shunted segment, the more sensitive is its profile to the balance of the effective rates $\alpha_2^*$ and $\beta_2^*$, consequently, the location of the shunted segment with coexisting low- and high-density phases is determined more precisely. The shift of the shortcut from that position upstream leads to a sharp change of the density profile resembling a transition to a high density phase, while the shift downstream causes a transition to a low density phase.

2. The main features of the coexistence phase and the density profiles in the whole network are well described by the domain wall theory. Apart from the negligible inter-chain correlations, they depend only on the current through the shortcut and do not depend on its structure.

3. The model displays an unexpected feature: the current through the longer shunted segment is larger than the current through the shortcut.

4. From the viewpoint of vehicular traffic, most comfortable conditions for the drivers are provided when the shortcut is shifted downstream from the position of coexistence, since then both the shunted segment and the shortcut exhibit low-density lamellar flow. Most unfavorable is the opposite case of upstream shifted shortcut, when both the shunted segment and the shortcut are in a high-density phase describing congested traffic of slowly moving cars.

It seems that the domain wall theory is applicable to the whole network because of the smallness of the inter-chain correlations generated by the random sequential update. Hence, it would be instructive to make a similar study of the steady state properties of TASEP on the same network but with particles obeying a stochastic dynamics in discrete time, e.g., the parallel update.

\section*{Acknowledgement} The partial support by a grant of the Plenipotentiary Representative of the Bulgarian Government at the Joint Institute for Nuclear Research, Dubna, is gratefully acknowledged.


\begin{thebibliography}{99}

\bibitem{MGP68} C.T. MacDonald, J.H. Gibbs, and A.C. Pipkin. \emph{Biopolymers} \textbf{6} (1968) 1.
\bibitem{D98} B. Derrida. \emph{Phys. Rep.} \textbf{301} (1998) 65.
\bibitem{CSS00} D. Chowdhury, L. Santen and A. Schadschneider. \emph{Phys. Rep.} \textbf{329} (2000) 199.
\bibitem{S01} G.M. Sch\"{u}tz. In: \emph{Phase Transitions and Critical Phenomena} vol. 19, edited by 
C. Domb and J.L. Lebowitz (London: Academic Press, 2000) pp 1-251.
\bibitem{H01} D. Helbing. \emph{Rev. Mod. Phys.} \textbf{73} (2001) 1067.
\bibitem{CMZ11} T. Chou, K. Mallick, and R.K.P. Zia. \emph{Rep. Prog. Phys.} \textbf{74} (2011) 116601.
\bibitem{NKP13}I. Neri, N. Kern, and A. Parmeggiani. \emph{New J. Phys.} \textbf{15} (2013) 085005.
\bibitem{NS92} K. Nagel and M. Schreckenberg. \emph{J. Phys. I} \textbf{2} (1992) 2221.
\bibitem{SSNI95}M. Schreckenberg, A. Schadschneider, K. Nagel, and N. Ito. \emph{Phys. Rev. E} \textbf{51} (1995) 2939.
\bibitem{DPPP12}A.E. Derbyshev, S.S. Poghosyan, A.M. Povolotsky, and V.B. Priezzhev. \emph{J. Stat. Mech.} (2012)    P05014.
\bibitem{PSSS01}V. Popkov, L. Santen, A. Schadschneider, and G.M. Sch\"utz. \emph{J. Phys. A} \textbf{34} (2001)  L45.
\bibitem{PR02}M.M. Pedersen and P.T. Ruhoff. \emph{Phys. Rev. E} \textbf{65} (2002) 056705.
\bibitem{JWW02}R. Jiang, Q.-S. Wu, and B.-H. Wang. \emph{Phys. Rev. E} \textbf{66} (2002) 036104.
\bibitem{BF08}S. Belbasi and M.E. Fouladvand. \emph{J. Stat. Mech.} (2008) P07021.
\bibitem{FSS04}M.E. Fouladvand, Z. Sadjadi, and M.R. Shaebani. \emph{Phys. Rev. E} \textbf{70} (2004) 046132.
\bibitem{Setal08}Y. Sugiyama, M. Fukui, M. Kikuchi, K. Hasebe, A. Nakayama, K. Nishinari, S.-I. Tadaki, and 
    S. Yukawa. \emph{New J. Phys.} \textbf{10} (2008) 033001.
\bibitem{BPB04} J. Brankov, N. Pesheva, and N. Bunzarova. \emph{Phys. Rev. E} \textbf{69} (2004) 066128.
\bibitem{PK05}E. Pronina and A.B. Kolomeisky. \emph{J. Stat. Mech.} (2005) P07010.
\bibitem{NKP11} I. Neri, N. Kern, and A. Parmeggiani. \emph{Phys. Rev. Lett.} \textbf{107} (2011) 068702.
\bibitem{WJNW09}X. Wang, R. Jiang, M.-B. Hu, K. Nishinari and Q.-S. Wu. \emph{Intern. J. Mod. Phys. C} \textbf{20} (2009) 1999.
\bibitem{SMJH11}X. Song, L. Ming-Zhe, W. Jian-Jun, and W. Hua. \emph{Chin. Phys. B} \textbf{20} (2011) 060509.
\bibitem{XTW11}S. Xia, L. Tang, and W. Wang. \emph{Cent. Eur. J. Phys.} \textbf{9} (2011) 1077.
\bibitem{ACH11}C. Appert-Rolland, J. Cividini, and H.J. Hilhorst. \emph{J. Stat. Mech.} (2011) P10014.
\bibitem{HA12}H.J. Hilhorst and C. Appert-Rolland. \emph{J. Stat. Mech.} (2012) P06009.
\bibitem{MSR12}L. Ming-Zhe, L. Shao-Da, and W. Rui-Li. \emph{Chinese Phys. B} \textbf{21} (2012) 090510.
\bibitem{PB13}N.C. Pesheva and J.G. Brankov. \emph{Phys. Rev. E} \textbf{87} (2013) 062116.
\bibitem{YJWHW07}Y.-M. Yuan, R. Jiang, R. Wang, M.-B. Hu, and Q.-S. Wu. \emph{J. Phys. A} \textbf{40} (2007) 12351.
\bibitem{BPB14}N. Bunzarova, N. Pesheva, and J. Brankov. \emph{Phys. Rev. E} \textbf{89} (2014) 032125.
\bibitem{KSN11}M. Kim, L. Santen, and J.D. Noh. \emph{J. Stat. Mech.} (2011) P04003.
\bibitem{Leduc12}C. Leduc, K. Padberg-Gehlec, V. Vargaa, D. Helbing, S. Dieza, and J. Howarda. \emph{PNAS} \textbf{106} (2012) 6100.
\bibitem{BB05}J. Brankov and N. Bunzarova. \emph{Phys. Rev. E} \textbf{71} (2005) 036130.
\bibitem{KSKS98}A.B. Kolomeisky, G.M. Sch\"{u}tz, E.B. Kolomeisky, and J.P. Straley. \emph{J. Phys. A} \textbf{31} (1998) 6911.
\bibitem{SA02}L. Santen and C. Appert. \emph{J. Stat. Phys.} \textbf{106} (2002) 187.

\end{thebibliography}
\end{document}